\documentclass[prl,twocolumn,superscriptaddress]{revtex4-2}   
\usepackage{graphicx,amssymb}
\usepackage{color,ulem}
\usepackage{bm}   
\usepackage{bm}
\usepackage{amssymb,amsmath}
\usepackage{amsthm}
\usepackage{latexsym}
\usepackage{graphicx,amssymb}
\usepackage{color,ulem}
\usepackage{float}
\usepackage{siunitx}
\usepackage{mathrsfs} 
\usepackage{soul}
 
\usepackage{graphicx}
\usepackage{dcolumn}
\usepackage{bm}
\usepackage[utf8]{inputenc}
\usepackage{amsfonts,amssymb,amsmath}
\usepackage[mathscr]{eucal}
\usepackage{epsfig}

\begin{document}
	
	\title{Bolometric detection of Josephson radiation}
	\author{Bayan Karimi}
	\affiliation{Pico group, QTF Centre of Excellence, Department of Applied Physics, Aalto University, P.O. Box 15100, FI-00076 Aalto, Finland}
	\affiliation{QTF Centre of Excellence, Department of Physics, Faculty of Science, University of Helsinki, FI-00014 Helsinki, Finland}
	\author{Gorm Ole Steffensen}
	\affiliation{Departamento de F\'{i}sica Te\'{o}rica de la Materia Condensada,\\Condensed Matter Physics Center (IFIMAC) and Instituto Nicol\'{a}s Cabrera,\\Universidad Autonoma de Madrid, 28049 Madrid, Spain}
	\affiliation{Instituto de Ciencia de Materiales de Madrid (ICMM), \\ Consejo Superior de Investigaciones Científicas (CSIC), \\ Sor Juana Inés de la Cruz 3, 28049 Madrid, Spain}
	\author{Andrew P. Higginbotham}
	\affiliation{The James Franck Institute and Department of Physics, University of Chicago, Chicago, IL 60637, USA}
	\affiliation{IST Austria, Am Campus 1, 3400 Klosterneuburg, Austria}
	\author{Charles M. Marcus}
	\affiliation{Materials Science and Engineering and Department of Physics, University of Washington, Seattle WA 98195}
		\affiliation{Center for Quantum Devices, Niels Bohr Institute, University of Copenhagen, 2100 Copenhagen, Denmark}
		\affiliation{InstituteQ – the Finnish Quantum Institute, Aalto University, Finland}
	\author{Alfredo Levy Yeyati}
	\affiliation{Departamento de F\'{i}sica Te\'{o}rica de la Materia Condensada,\\Condensed Matter Physics Center (IFIMAC) and Instituto Nicol\'{a}s Cabrera,\\Universidad Autonoma de Madrid, 28049 Madrid, Spain}
	\author{Jukka P. Pekola}
	\affiliation{Pico group, QTF Centre of Excellence, Department of Applied Physics, Aalto University, P.O. Box 15100, FI-00076 Aalto, Finland}

	\date{\today}
	
	\begin{abstract}
		A Josephson junction (JJ) has been under intensive study ever since 1960's. Yet even in the present era of building quantum information processing devices based on many JJs, open questions regarding a single junction remain unsolved, such as quantum phase transitions, coupling of the JJ to an environment and improving coherence of a superconducting qubit. Here we design and build an engineered on-chip reservoir that acts as an efficient bolometer for detecting the Josephson radiation under non-equilibrium (biased) conditions. The bolometer converts ac Josephson current at microwave frequencies, up to about $100\,$GHz, into a measurable dc temperature rise. The present experiment demonstrates an efficient, wide-band, thermal detection scheme of microwave photons and provides a sensitive detector of Josephson dynamics beyond the standard conductance measurements. Using a circuit model, we capture both the current-voltage characteristics and the measured power quantitatively.
	\end{abstract}
	\maketitle
	\begin{figure*}
		\centering
		\includegraphics [width=\textwidth] {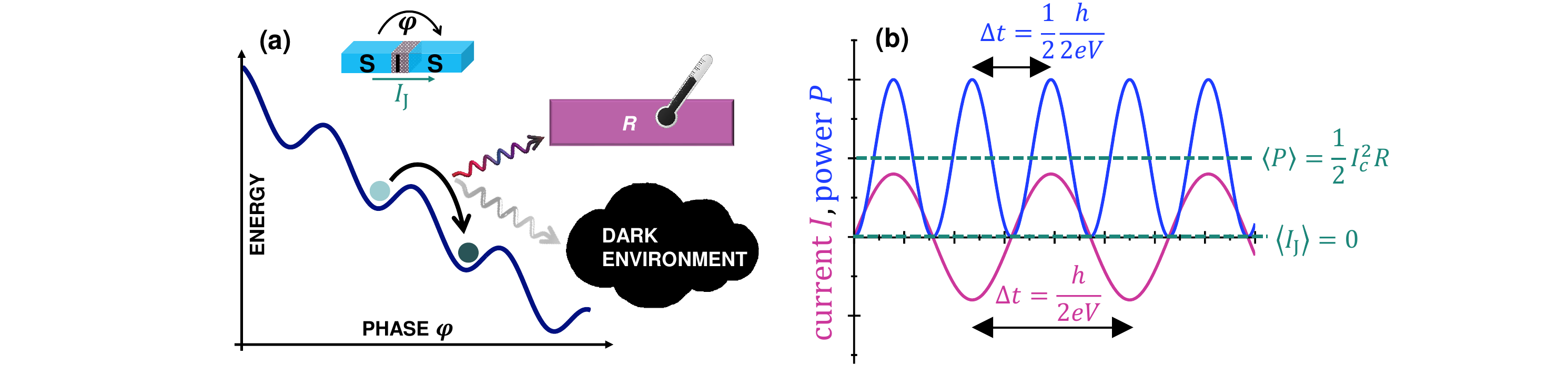}
		\caption{Energy released from a biased superconductor-insulator-superconductor (SIS) Josephson junction. (a) The conceptual illustration of the energy vs. the phase drop $\varphi$ across the biased JJ. This junction emits energy either to the engineered absorber on the chip $R$ or to the dark environment. (b) The time average current $\langle I_{\rm J}\rangle=0$ vanishes ideally in a voltage-biased junction, whereas the average power remains non-vanishing. 
			\label{fig1}}
	\end{figure*}
Understanding the dissipative dynamics of a Josephson junction (JJ)~\cite{Josephson62,Giaever,Tinkham,Devoret,Vion,Golubov} has been a topic of intensive studies ever since the seminal theoretical works of Ivanchenko and Zil'berman~\cite{iz69} and of Caldeira and Leggett~\cite{caldeira83}. Nowadays, a JJ serves as a versatile component with a wide range of applications in various areas such as quantum computing and metrology~\cite{tafuri19}. Radiation from different types of JJs including semiconductor nanowires and thin-film microbridges has been detected by observing photon assisted tunneling, current-phase relationship, and properties of resonant cavities~\cite{Clark,Lindelof,Geresdi,Haller,bretheau,Bretheau2,Deacon,Haller2}. The concept of our experiment – converting ac Josephson current at microwave frequencies to measurable dc power – is based on nano-bolometric techniques~\cite{Irwin,Qdetector,BK,Kokkoniemi,Hadfield,Walsh,phase-slip,Ibabe}. We employ a hot-electron bolometer (HEB), which consists of a normal-metal nano-absorber whose temperature is measured by a normal-metal – insulator – superconductor (NIS) thermometer probe~\cite{nahum93,RMP2006}. Due to the quadratic response of the bolometer, simple dc measurement of the absorber temperature yields the magnitude of the Josephson current at frequencies up to about 100 GHz. 

In the experiment, we simultaneously measure the average (charge) current through the junction and the HEB signal. The temperature of the HEB is able to resolve the transport characteristics of the Josephson junction with high resolution, and it reveals complementary features as compared to those given by the dc charge transport characteristics. As a byproduct, we find that the ubiquitous drop of current at finite subgap voltage can be quantitatively explained by the shunting effect of the local environment of the junction.
	
	\begin{figure*}
		\centering
		\includegraphics [width=\textwidth] {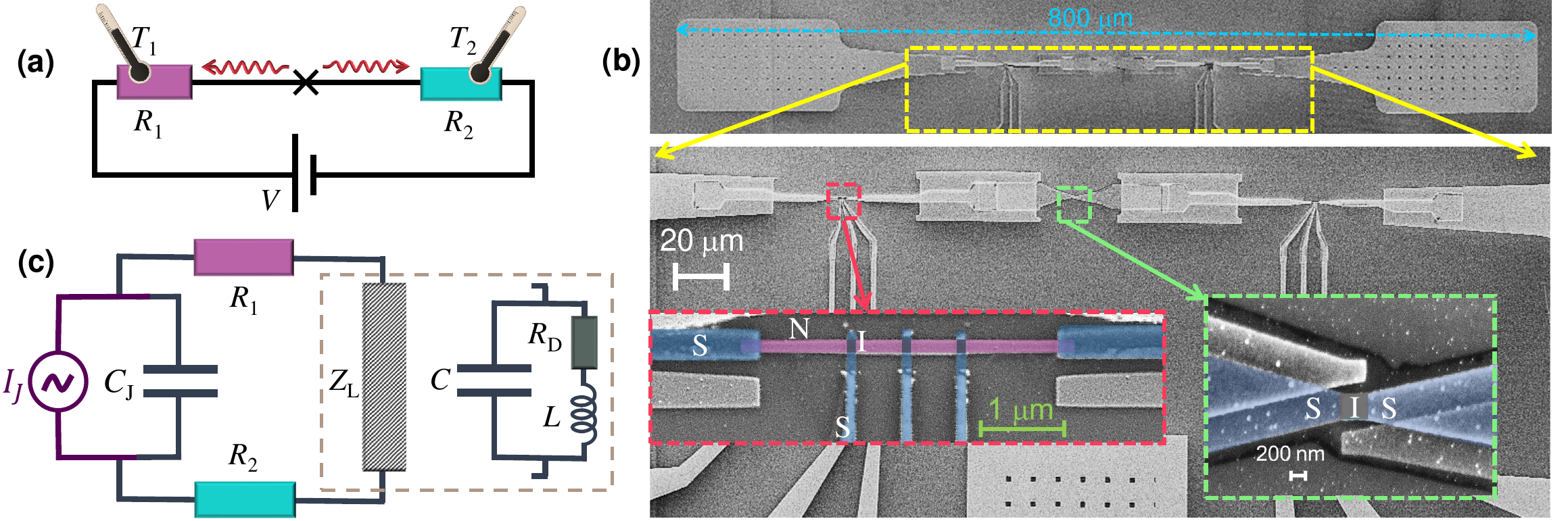}
		\caption{A setup consisting of the JJ surrounded by two bolometers. (a) A schematic illustration of the device where the energy from the voltage $V$-biased JJ (cross shaped structure) at angular frequency $\omega_{\rm J}=2eV_{\rm JJ}/\hbar$ is emitted to the two absorbers with resistances $R_1$ and $R_2$. Here $V_{\rm JJ}\simeq V$ is the voltage drop across the JJ. The thermometers on the absorbers measure the temperature changes of them, $T_1$ and $T_2$, simultaneously. (b) A scanning electron micrograph of the device on different scales. A single JJ in the middle is sandwiched between two resistors placed at a distance of about 100~$\mu$m from the JJ. The connections are made with aluminum (Al) and with niobium (Nb, the wider sections). The zoomed-in views highlight the SIS junction (Al, blue; AlO$_x$, grey) in the middle (scale bar, 200~nm) and the left side, one of the absorbers (resistor $R_1$) made of copper in purple color, in clean contact with Al (blue) leads (scale bar, 1~$\mu$m) connecting to the patterned niobium film (light grey). The temperature of the absorber is monitored and controlled by three NIS probes (Cu/AlO$_x$/Al). (c) Lumped-element model of the device for radio-frequencies; a single JJ represented by a  current source $I_{\rm J}$, an ideal Josephson element, and capacitance $C_{\rm J}$ in parallel. The load impedance $Z_{\rm L}$ is the termination of the device via bonding pads and wire bonds, and it is modelled as a dissipaive $LCR_{\rm D}$ element as shown. Here $R_{\rm D}$ is the resistor representing the dark environment.
			\label{fig2}}
	\end{figure*}
	
	The conventional picture of a current-biased Josephson junction (JJ) is that of a tilted washboard potential against the phase bias $\varphi$ across it as in Fig.\,\ref{fig1}(a), where the slope is given by the bias. At small currents, the phase remains trapped in one of the wells of this potential, and since the voltage $V$ is proportional to the time derivative of this phase, given by the Josephson relation as  $V=\frac{\hbar}{2e} d\varphi/dt$, the junction remains in its zero-voltage supercurrent state. The current in the junction is thus dissipationless. Upon increasing the tilt of the potential, the phase starts to move from one well to the next lower one, producing a non-vanishing voltage, i.e. the process becomes dissipative. Each jump in phase from one well to the next lower one releases energy equal to $\Phi_0 I$, where $\Phi_0=h/2e$ is the superconducting flux quantum and $I$ is the average current. The current experiment deals, however, mainly with a voltage-biased configuration where the picture is less understood.
	
	The current through the junction at non-vanishing voltage is given by the Josephson relation, $I_{\rm J} =I_c \sin \varphi$, where $I_c$ is the junction-specific critical current, and the phase is evolving according to $\varphi (t) =2eVt/\hbar$. The mean-squared value of this current is then given by $I_c^2/2$. The conceptual difference between the measurement of current and power is illustrated in Fig.\,\ref{fig1}(b). A key question is, can this alternating ac Josephson current be fully detected by the bolometer operating in dc mode, i.e., can we measure the power equal to $P=I_c^2 R/2$ via a temperature measurement of the resistor with resistance $R$? We will first introduce the experimental setup and results, which allow us to answer this question affirmatively. 
	The second key question in our work is: where does this energy get absorbed? Can it be collected by a bolometric detector or is it released to an unknown environment which can not be monitored? 
		\begin{figure*}
		\centering
		\includegraphics [width=\textwidth] {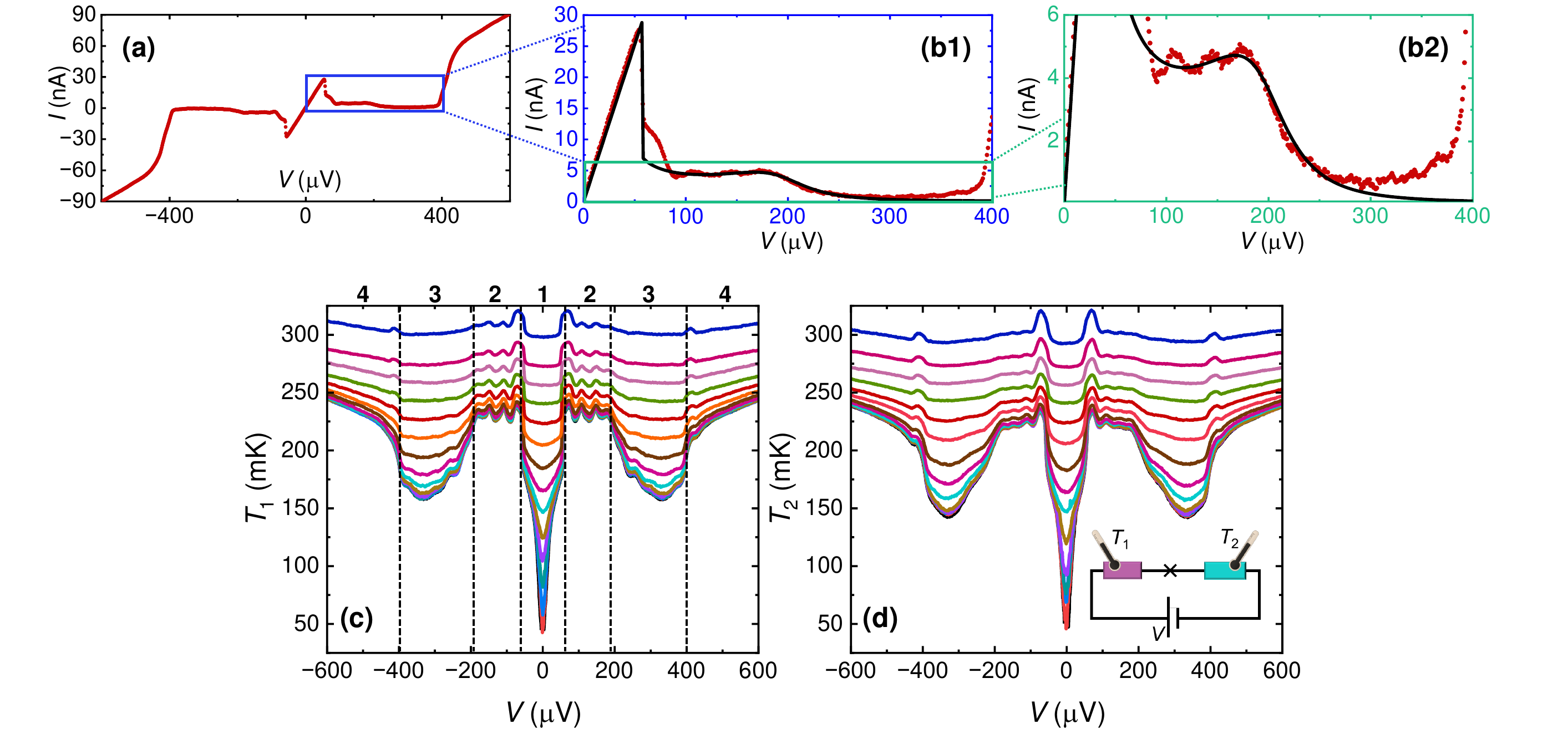}
		\caption{Transport characteristics of the device of Fig.\,\ref{fig2}(b). (a) The {\it I-V} characteristic shown with red symbols measured at $T=50$~mK. 
			(b1),(b2) Zoom-in of the enclosed area in blue shown in panel (a). The black solid line is from the theoretical model with parameters: $R_1+R_2=30\,\Omega$, $R_{\rm D}=120\,\Omega$, $R_{\rm S}=1920\,\Omega$, $C_{\rm J}=15\,$fF, $C=40\,$fF, $L=2.1\,$nH, and $I_c=64\,$nA. (c),(d) Simultaneous measurement of the temperature  of the two resistors ($T_1$ and $T_2$) as a function of applied bias $V$. 
			\label{fig3}}
	\end{figure*}
	
A schematic representation of the setup is shown in Fig.\,\ref{fig2}(a) where the energy is released by the JJ biased at voltage $V$, indicated by the cross. As will be clear in what follows, the bias voltage $V$ is dropping almost fully across the junction except in the trivial supercurrent branch. Therefore we do not make a distinction between the two, and use $V$ to denote both. The temperatures $T_1$ and $T_2$, of the absorbers are measured by the thermometers attached to them. Figure\,\ref{fig2}(b) displays the actual device, where a single JJ in the middle is positioned between two normal metal resistors $R_1$ and $R_2$, made of copper, working as absorbers, at a distance of about $100\,\mu$m symmetrically around the JJ. The connections are made of aluminum whose thermal conductance is negligible at low temperatures \cite{Joonas2010, RMPJB}. By current-biasing a pair of normal-metal – insulator – superconductor junctions (SINIS) attached by superconducting leads to the resistors, one can control and monitor their temperatures. The voltage across the junction at a fixed bias current measured at different bath temperatures yields the temperature calibration in a standard way as presented in the Supplementary Material. The resistance of each resistor is about 15 $\Omega$. The simplified equivalent circuit of the device, used in the theoretical modeling of the {\it I-V} characteristics and power, is presented in Fig.\,\ref{fig2}(c). Here a single JJ is represented by an ac current source in parallel with capacitance $C_{\rm J}$, and $Z_{\rm L}$ is the external impedance formed mainly by the bonding pads and wires.
	
				\begin{table}[h]  
	\caption{Different bias regimes} 
	\centering 
	\begin{tabular}{l c c c c} 
		\hline\hline   
		Regime &Range of $|V|$ &Process  
		\\ [0.5ex]  
		\hline   
		& & \\[-1ex]  
		{Regime 1} & {$<50\,\mu$V}&supercurrent
		\\[1ex]  
		& &Josephson radiation I \\[-1ex]  
		\raisebox{1ex}{Regime 2} & \raisebox{1.5ex}{$50-200\,\mu$V}& $\omega<\omega_{\rm LC}$
		\\[1ex]  
		& &Josephson radiation II  \\[-1ex]  
		\raisebox{1ex}{Regime 3} & \raisebox{1.5ex}{$200-400\,\mu$V}& $\omega>\omega_{\rm LC}$  
		\\[1ex]  
		& &  \\[-1ex]  
		{Regime 4} & {$> 400\,\mu$V}& quasiparticle current
		\\[1ex]
		\hline 
	\end{tabular}  
\end{table} 
	The {\it I-V} characteristic of the device at $T=50$~mK is presented in Fig.\,\ref{fig3}(a). In the supercurrent branch in the central part, the positive constant slope is a result of the series resistance $R_{\rm S}$ from the two absorbers, the load resistors on the chip, and line resistance from room temperature to the chip ($R_{\rm S}\approx 1920~\Omega$) as we do a two-wire measurement. The almost flat section up to the gap, $V=2\Delta/e\simeq 400\,\mu$V, is followed by a rise in current, transitioning into a quasiparticle current with specific resistance of the JJ, $R_T=5.0\,$~k$\Omega$. The magnified view of the enclosed area of the {\it I-V} characteristic in panel (a), as displayed in Figs.\,\ref{fig3}(b1) and \ref{fig3}(b2), provides a closer look at this measurement. It unveils additional details, particularly the oscillations between $|V|=50-200\,\mu$V and drop in current at $|V|=\Delta/e\simeq 200\,\mu$V in the subgap regime.
  \begin{figure*}
	\centering
	\includegraphics [width=\textwidth] {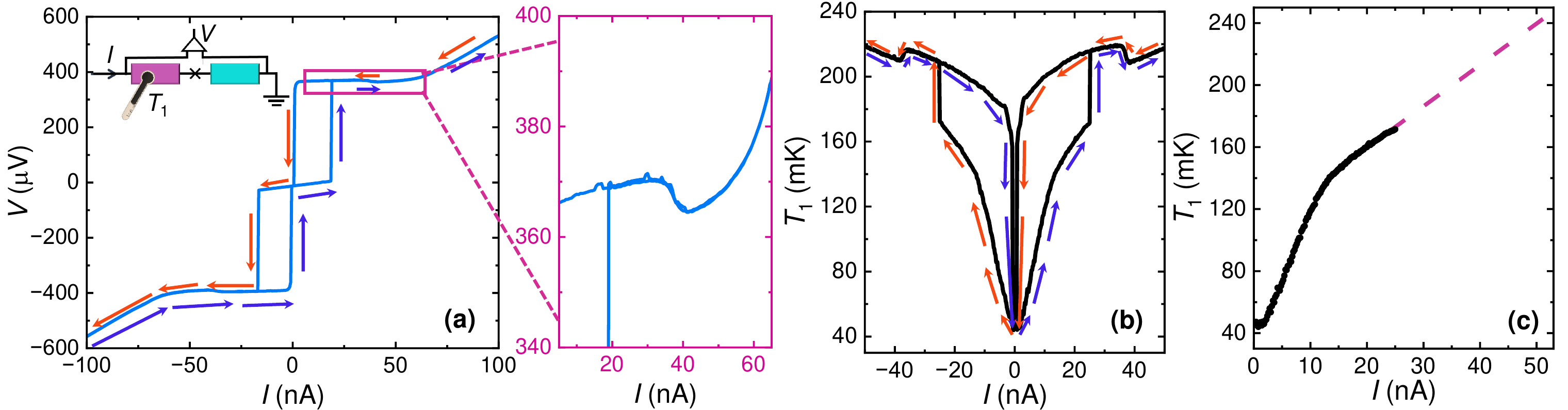}
	\caption{ Characteristics of the current-biased JJ. (a) Hysteretic $V-I$ characteristics of the junction together with a zoom in the quasiparticle branch. The blue arrows display the signal in the forward current sweep while the red ones present the reverse sweep. The setup for voltage and temperature measurement is shown in the inset. (b) Measured resistor temperature at current bias of $I_{1,{\rm th}}=15$~pA at $T_0=43~$mK. The nearly parabolic bottom in the small bias range corresponds to the hysteretic supercurrent branch of the single JJ. (c) The data in the supercurrent branch presented in panel (b) at positive currents. The dashed line indicates the linear extrapolation in the regime not reachable by the supercurrent measurement. 
		\label{fig5}}
\end{figure*}
	
	We now analyze theoretically the {\it I-V} profile in the almost flat subgap region, $50\,\mu{\rm V}\lesssim V \lesssim 400\,\mu{\rm V}$, in terms of an underdamped JJ in series with a frequency dependent impedance $Z(\omega)$. In this regime, the phase evolves as
		\begin{equation}
			\varphi(t) \approx \omega_{\rm J}t + \frac{2e}{h}\frac{I_c |Z(\omega_{\rm J})|}{\omega_{\rm J}}\sin\left(\omega_{\rm J} t+\delta\right)
		\end{equation}
		with the phase shift $\delta = -\arctan(\text{Re}Z(\omega_{\rm J})/\text{Im}Z(\omega_{\rm J}))$. To the lowest order in voltage variations across the junction, the JJ dc current $I\equiv \langle I_{\rm J}\rangle$ and voltage drop are given by,
		\begin{equation}\label{eq:CurrVolt}
			I=I = I_c^2 \frac{e}{h} \frac{\text{Re}Z(\omega_J)}{\omega_J}, 
			~~V= \frac{h}{2e} \omega_{\rm J} + Z(0)I,
		\end{equation}
		which is solved to yield the {\it I-V}. Intuitively, the proportionality of $I$ on $Z(\omega)$ arises from the dynamics of the phase particle in the washboard potential. The lower the friction, given by $Z(\omega)$, more uniformly the phase runs down the washboard resulting ideally in vanishing $I = I_c \langle \sin\varphi(t)\rangle_0$, while larger friction results in a more rugged motion of it, thereby increasing $I$. Here the subscript 0 denotes the average over the period. The {\it I-V} seen in Figs.~\ref{fig3}(b) is reproduced by using the impedance of the ac circuit, containing a $LC$ resonance, $\omega_{LC}/2\pi\approx 100$~GHz. More details on the theory can be found in the Supplementary Material. This model captures the main features of the experimental {\it I-V} curve with realistic circuit parameters as shown by black lines in Figs.\,\ref{fig3}(b1) and \ref{fig3}(b2).
	
	The energy released from the JJ at the given voltage bias is absorbed partly or fully by the two resistors. The measurement of temperatures $T_1$ and $T_2$ as a function of applied bias $V$ is presented in Figs.\,\ref{fig3}(c) and \ref{fig3}(d), respectively, see also additional data for another sample in Fig.~\ref{fig3-add}. The temperature calibration is explained in the Supplementary Material. The non-monotonic behavior of temperatures is remarkably similar in the two absorbers. The temperature changes clearly indicate four different regimes based on the different sections of the {\it I-V} characteristic presented in panels \ref{fig3}(a) and \ref{fig3}(b). These regimes are summarized in Table 1. In the central region, Regime 1 up to $|V|\simeq 50\,\mu$V, the entire heating of the resistors is attributed to Joule heating, a consequence of the current flowing through the resistors, as there is no power dissipation within the Josephson junction itself. The most interesting regime, Regime 2, lies within the intermediate range, $50\,\mu{\rm V}< |V|< 200\,\mu{\rm V}$, where the energy released by the ac Josephson current at the frequency $f=2eV/h$ is absorbed by the resistors. In this regime, the temperature is high and shows small oscillations as a function of $I$, as discussed below and in the additional data of Fig.\,\ref{fig4}. In Regime 3, $200\,\mu{\rm V}< |V| < 400\,\mu{\rm V}$, the power emitted into the two resistors decreases, corresponding to reduction in their temperatures. At high voltage bias $|V|>400\,\mu$V, Regime 4, in the quasiparticle branch, only a small fraction of power generated within the junction is dissipated into the two absorbers leading to a continuous, monotonous increase in their temperatures. Note that power generated at the junction in Regime 4, based on the dc {\it I-V}, is indeed more than 100 times higher than in Regime 2, but the temperature rise is comparable indicating a different mechanism of heat transport in each regime. 
	
	\begin{figure*}
		\centering
		\includegraphics [width=\textwidth] {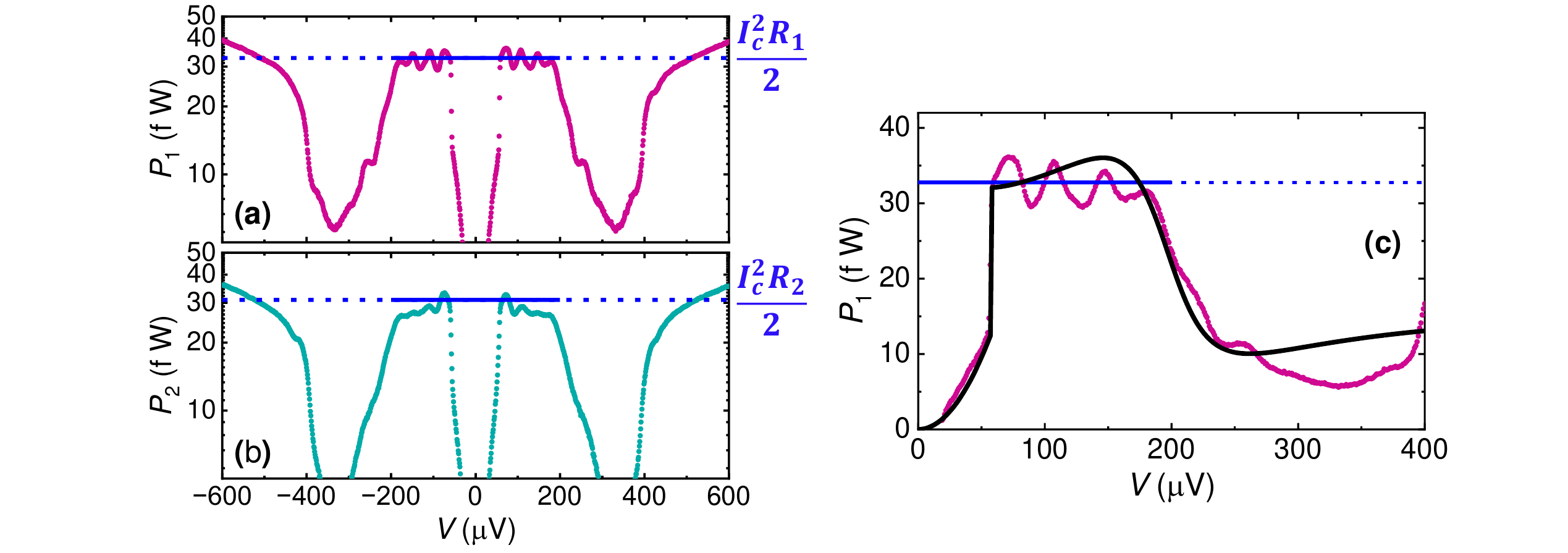}
		\caption{The power $P_i$ for $i=1,\,2$ as a function of the applied voltage $V$, obtained from the temperature measurement $T_i$ for the two bolometers (Figs.\,\ref{fig3}(c) and \ref{fig3}(d)) at $T_0=43$~mK are shown in panels (a) and (b). The conversion from $T$ to $I$ is obtained from the calibration presented in Fig.\,\ref{fig5}(c). The horizontal blue line in each panel corresponds to $I_c^2R_i/2$. (c) In this panel we have included the theoretical model, Eq.\,(3), to (a) using the same parameter values as for the {\it I-V} characteristic measurement presented in Fig.~\ref{fig3}.
			\label{fig6}}
	\end{figure*}
	
	The oscillations in Regime 2, spaced by about $\Delta V=50\,\mu$V, seen in Figs.\,\ref{fig3}(c) and \ref{fig3}(d) and expanded in the additional data in Fig.\,\ref{fig4}, can be qualitatively understood as follows. The whole device is about $800\,\mu$m long as shown in the top of Fig. \ref{fig2}(b). It acts as a cavity with imperfect termination. This termination by bonding pads and inductive bonding wires causes partial reflection of current leading to temperature variations in the absorbers visible in the subgap regime. The inset of the additional data in Fig.\,\ref{fig4} displays the temperature variation of the absorber 1 within the subgap regime. The blue arrows point to the minima of the measured temperature. The main panel of this figure shows the positions of the minima in both absorbers representing perfect coincidence of the two.
	
	Our qualitative interpretation of the overall temperature characteristics in Figs.\,\ref{fig3}(c) and \ref{fig3}(d) and in the inset of Fig.~\ref{fig4} is as follows. Beyond the pure Joule heating of the resistors by the dc current in Regime 1, the high level of heating in Regime 2 is attributed to the absorption of the ac-supercurrent-induced high-frequency ($2eV/h$) power in the corresponding resistor. As shown below, this dc power is very close to the average value $P_i=\frac{I_c^2}{2}R_i$ for $i=1,\,2$. Crossover to Regime 3 is characterized by an abrupt decrease of power for both resistors. This drop has been seen in all the samples that we have measured, and it can be accounted by the drop in the current $I$: the circuit re-routes the current away from the thermometers at higher frequencies. 
 
To understand this mechanism of re-routing we note that based on our circuit model (Fig.~\ref{fig2}(c)), the power deposited in the HEBs in the underdamped regime is given by,
		\begin{align}
			P_{i} &= R_{i}I^2 \\ 
			&+ \frac{R_i I_c^2}{2}\left[\left(1+C_{\rm J}\omega_{\rm J} \text{Im}Z(\omega_{\rm J})\right)^2 + \left(C_{\rm J} \omega_{\rm J} \text{Re}Z(\omega_{\rm J})\right)^2\right],
			\nonumber
		\end{align}
		which for small biases results in $P_I \approx \frac{R_i I_c^2}{2}$, but for biases $V\gg \frac{h}{2e}\omega_{LC}$ the power approaches $P_i \approx\frac{R_iI_c^2}{2}\frac{1}{(1+C_{\rm J}/C)^2}$. This is a consequence of the ac current bypassing the HEBs by running through $C$ and $C_{\rm J}$, instead of $L$, at large frequencies.
	Regime 4 represents pure quasiparticle current in tunneling, where the heat is released to the electrodes right at the junction in form of hot quasiparticles. These non-equilibrium quasiparticles diffuse poorly in the superconductor and do not carry heat to the resistors as effectively as Josephson radiation does. 
	
The hysteretic measurement in the current biased configuration, presented in Fig.\,\ref{fig5}(a) gives us a way to perform the power calibration for the HEB. The key to this calibration is the measurement of the rise of the temperatures of the resistors $R_i$ in the supercurrent branch of the Josephson junction, where there is no dissipation in the junction, and all the power is the plain Joule heating by the dc current through the corresponding resistor. The data from one of them as a function of current $I$ is shown at $T_0=43$ mK in Fig.\,\ref{fig5}(b). The blue arrows mark the current sweep in the forward direction, whereas the red ones denote the reverse direction. The hysteretic range at low currents corresponds to the non-dissipative supercurrent branch of the single JJ. Interestingly, we also see a drop in temperature at about 35 nA in the non-hysteretic quasiparticle branch. The $V-I$ shown in Fig.\,\ref{fig5}(b) and its inset demonstrates a drop at the same value of current, consistent with reduced heating above 35 nA. This drop could be associated with the “back-bending” behavior that has been observed in superconducting tunnel junctions~\cite{Yeh,Winkler} due to a suppression of the gap caused by the non-equilibrium distribution of the tunneling quasiparticles. 

In Fig.\,\ref{fig5}(c), we present the measurement of temperature for positive currents only. To reach the temperatures in all the relevant ranges, we perform a linear extrapolation. Now having the relation between $I$ and $T_i$, we can convert each measured temperature $T_i$ to effective power also in the voltage biased configuration. The above calibration allows us to convert temperature plots presented in Figs.\,\ref{fig3}(c) and \ref{fig3}(d) to power injected to the resistor, $P_i=I^2 R_i$ for $i=1,\,2$, as a function of bias voltage as presented in Figs.\,\ref{fig6}(a) and \ref{fig6}(b). The maximum power injected in the subgap regime is about $35\,$fW in each resistor. This power corresponds to about 25\% of $IV$ at low bias voltages.
	
	The single JJ in the voltage biased configuration produces an ac Josephson current. We may compare the measured power to that expected if the ac Josephson current would produce Joule heating of resistor $i$ at the rate $P_i=I_c^2 R_i/2$, stipulated by the condition that all this current would pass through this purely resistive absorber. The copper wire can indeed be considered as a resistive element as its inductance and the capacitance across are small enough at the relevant microwave frequencies ($< 100$ GHz). The critical current can be obtained from the Ambegaokar-Baratoff formula \cite{Ambegaokar-Baratoff}, $I_c=\pi \Delta/(2eR_T)$, applicable at temperatures far below the critical temperature $T_c \approx$ 1.3 K as in our experiment. The junction parameters were determined from the {\it I-V} curve of the JJ, with values $\Delta/e = 210\,\mu$V and $R_T=5.0\,{\rm k\Omega}$, yielding $I_c = 64\,$nA. With this procedure we obtain the expected powers $P_1=37\,$fW and $P_2=35\,$fW, indicated by the blue horizontal lines in Fig.\,\ref{fig6}, in good agreement with the maximum sub-gap power in the experiment. The result of the theoretical model is shown by the solid black line in Fig.~\ref{fig6}(c). Using the same parameter values as in Fig.~\ref{fig3}, the full theoretical model captures the main features of the bias dependence in the experiment.
	
	In this article, we have demonstrated experimentally bolometric detection of ac Josephson current. It is a complementary dc method with respect to measuring the {\it I-V} characteristics: the standard {\it I-V} or conductance measurement essentially averages the sinusoidal current, whereas the HEB measurement is quadratic in current, thus rectifying and fully retaining the ac component. This way, still by a simple dc measurement, we gain access to fine details of ac characteristics, which are largely averaged out and thus unreachable in the standard transport measurements. We presented a theoretical model that agrees with the experimental results using realistic circuit parameter values. It also gives a way to estimate the value of $R_{\rm D}$, which presents the unknown dark environment. Our model also shows that one could improve the detection efficiency of the HEB by increasing its resistance as this quantity scales approximately as $(R_1+R_2)/(R_{\rm D}+R_1+R_2)$.
\section{Acknowledgments} We thank Mikko Möttönen, Diego Subero, Vasilii Vadimov, Arman Alizadeh, Christoph Strunk, Nicolas Roch, Sergey Kafanov, Sergey Kubatkin, Andrew Kerman, and Joonas Peltonen for scientific discussions and Ze-Yan Chen for technical assistance. This work was funded by the Research Council of Finland Centre of Excellence program grant 336810 and grants 349601 (THEPOW). G. S. and A. L. Y. acknowledge financial support from the Spanish Ministry of Science through Grant TED2021-130292B-C43 funded by MCIN/AEI/10.13039/501100011033, "ERDF A way of making Europe" and the EU through FET-Open project AndQC. A. H. acknowledges support from the NOMIS foundation. C. M. M. acknowledges support from the Danish National Research Foundation and a research grant (Project 43951) from VILLUM FONDEN. We thank the facilities and technical support of Otaniemi research infrastructure for Micro and Nanotechnologies (OtaNano).

	\section*{Methods} 
	{\bf Fabrication:} Our process consists of three steps: fabricating (i) the Nb patterned ground planes, (ii) single SIS JJ, and (iii) absorbers and thermometers. The devices were fabricated on highly resistive $675\,\mu$m thick silicon substrates onto which a $40\,$nm aluminum-oxide layer by atomic layer deposition has been grown. Next it is coated with $200$-nm-thick sputtered niobium film. Broader features such as contact leads and bonding pads were patterned by reactive ion etching (RIE) using CF6+Ar chemistry on an electron-beam lithography-defined mask. The Josephson junction and NIS tunnel junctions were fabricated using a two-step process. First, shadow-mask electron-beam lithography was employed to pattern them onto a $950$-nm-thick poly(methyl-metacrylate)/copolymer resist bilayer. This step was followed by developing the exposed structures in methyl-isobutyl-ketone (MIBK) in isopropanol (IPA) developer and methylglycol-methanol solution for making undercuts, respectively. 
	
	Next, the junctions were deposited using the standard Dolan bridge technique by an electron-beam evaporator. Prior to metal evaporations, {\it in-situ} argon plasma milling was conducted on the sample surface. This milling process aimed to remove the native oxide layer, providing a pristine interface between the Nb and Al layers. In the first step, two-angle deposition process of $30$-nm-thick Al layers with intermediate {\it in-situ} oxidation are performed to realize the single Josephson junction with tunnel resistance of $5\,{\rm k\Omega}$ at mK temperatures. 
	
	The bolometers are made in the second step as shown in Fig.\,\ref{fig2}, similar to the preceding step. The procedure initiates with {\it in-situ} Ar ion plasma milling to ensure a clean contact between Nb and the deposited metals. The procedure then continues by first depositing a $20$-nm-thick Al layer and {\it in-situ} oxidation, followed by a $30$-nm-thick Cu layer, and finally by a $50$-nm-thick Al layer in clean contact with the Cu layer. The typical tunnel junction resistance for NIS for this experiment is about $20\,{\rm k\Omega}$. These NIS electrodes (probes) are connected to bonding pads for the setting and readout of the electronic temperature of the absorbers ($R_1$ and $R_2$). 
	
	The ﬁnal stage of the fabrication is lift-off in acetone ($52$ degrees for $\sim\,20-30\,$min) and cleaning in isopropyl alcohol. After the fabrication process the device is bonded with Al wires to a custom-made brass chip carrier for the cryogenic characterization.

	\section*{Additional information}
	\begin{figure}
		\centering
		\includegraphics [width=\columnwidth] {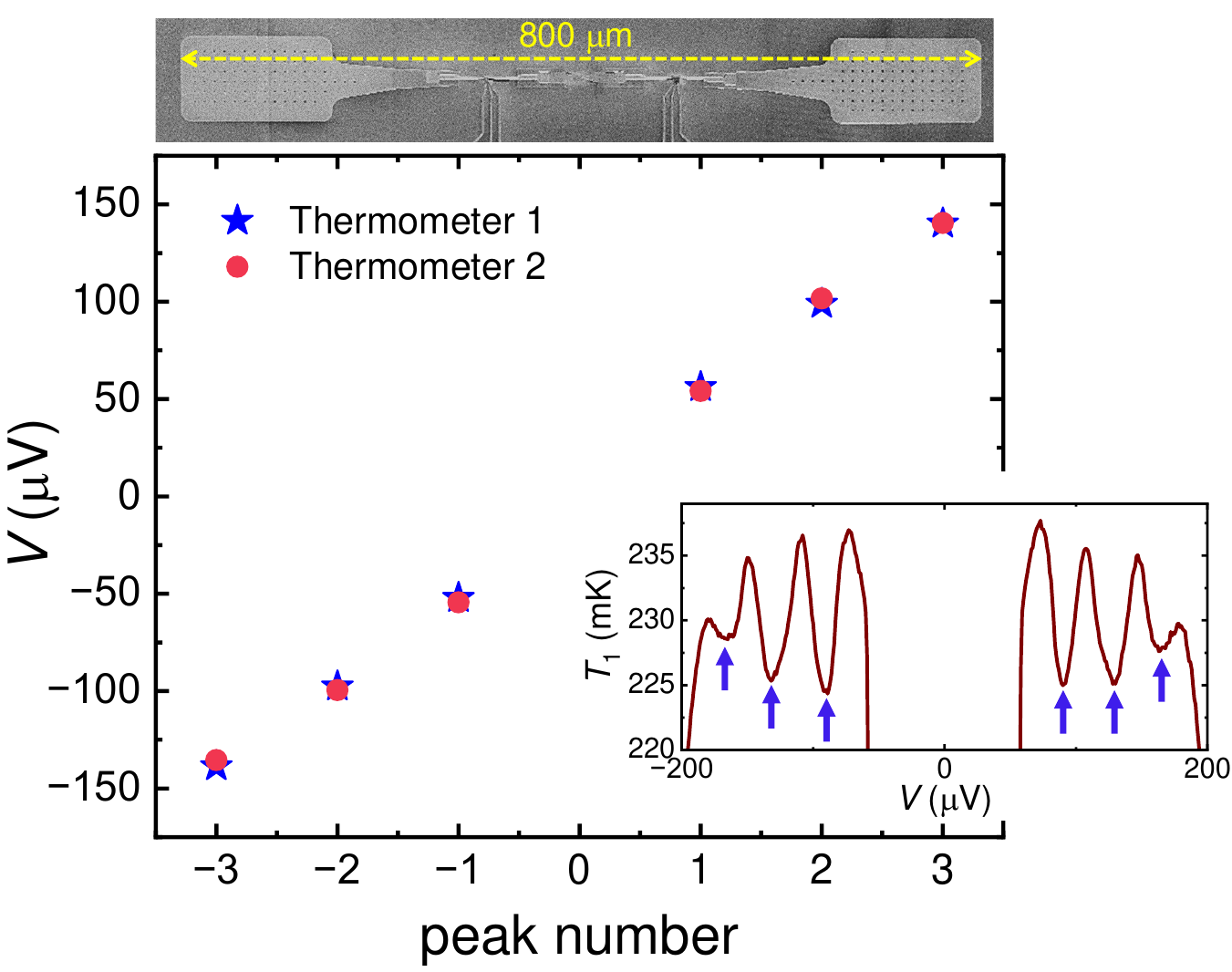}
		\caption{Temperature resonances in the JJ cavity. Extracted voltage bias points $V$ versus the peak number of both bolometers presented in Figs.\,\ref{fig3}(c) and \ref{fig3}(d). The inset shows a magnified view of the bias dependent temperature of $R_1$ at base temperature $T_0=43$~mK. The arrows point to the minima in temperature that have been extracted to the main panel.
			\label{fig4}}
	\end{figure}

	\begin{figure*}
		\centering
		\includegraphics [width=\textwidth] {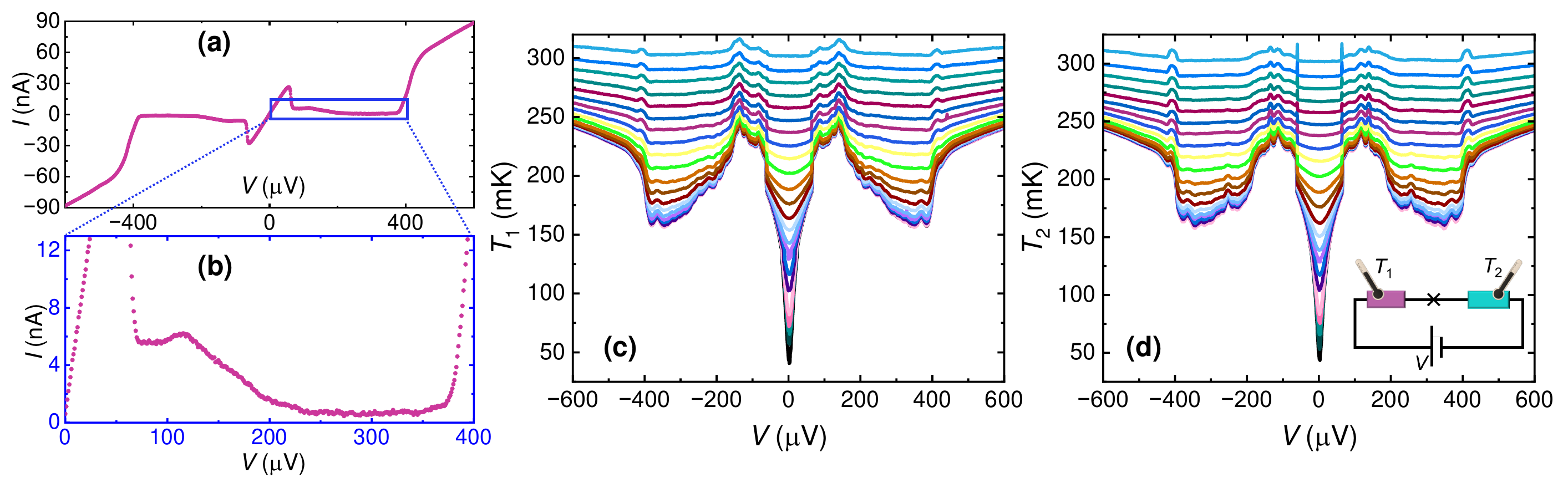}
		\caption{Similar data as in Fig.~\ref{fig3}, but here measured on another sample with nominally same parameters as those of the sample in the main text but with different wire-bonding conditions. 
			\label{fig3-add}}
	\end{figure*}

\cleardoublepage 

\appendix

\counterwithin*{figure}{part}

\stepcounter{part}

\renewcommand{\thefigure}{A.\arabic{figure}}

		\section{Supplementary Material}

	\section*{Temperature calibration}
\begin{figure*}[h!]
		\centering
		\includegraphics [width=\textwidth] {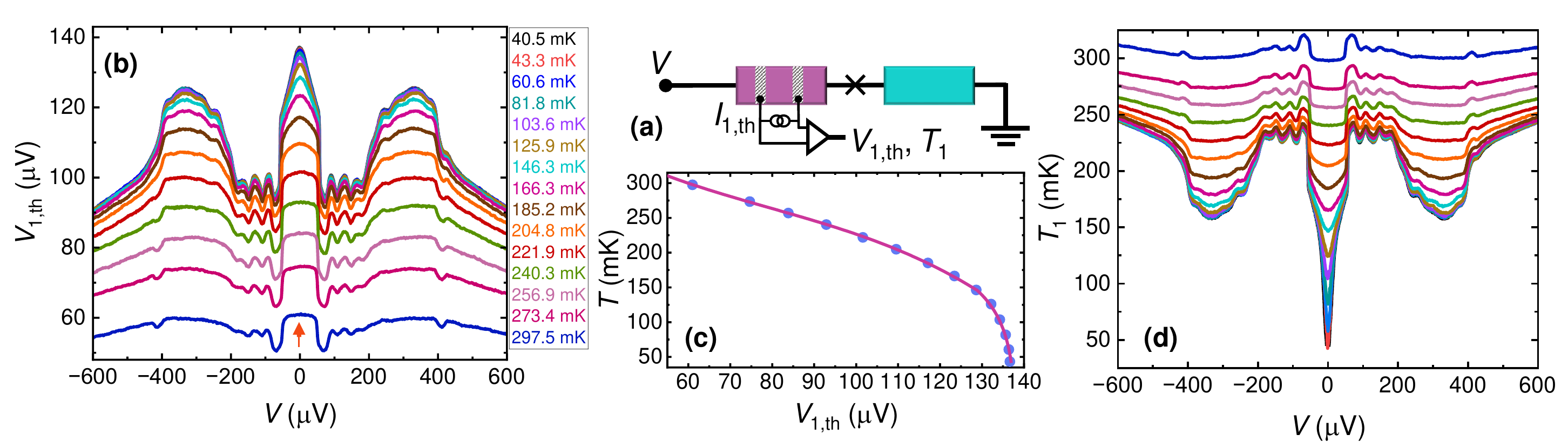}
		\caption{Temperature calibration of the voltage-biased JJ. (a) Current-biased measurement setup. The thermometer voltage $V_{\rm 1,th}$ is measured at a fixed current $I_{\rm 1,th}=15\,$pA as shown in (a) at different bath temperatures $T_0=43-298\,$mK from top to bottom, against the voltage across the JJ. (d) The equilibrium values of $V_{\rm 1,th}$, at $V=0$ (the position is shown by the red arrow in panel (c), serves as the calibration $V_{\rm 1,th} (V=0)$, versus $T$.}
		\label{fig.voltage-biased-meas}
	\end{figure*}

\begin{figure*}[h!]
		\centering
		\includegraphics [width=\textwidth] {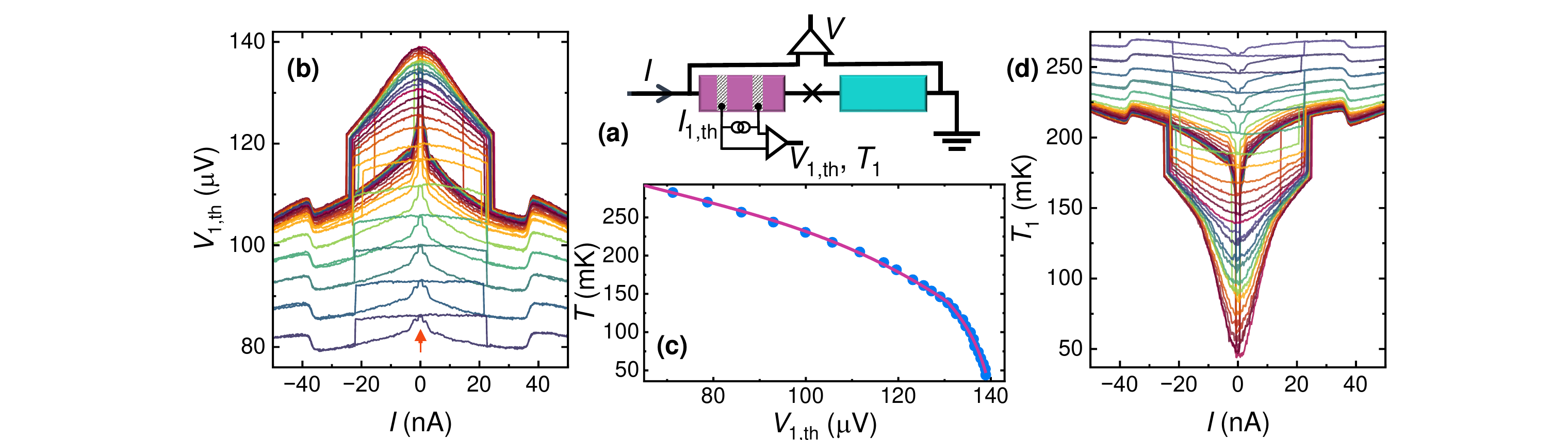}
		\caption{Temperature calibration of the current-biased JJ. (a) Current-biased measurement setup. The thermometer voltage $V_{\rm 1,th}$ is measured at a fixed current $I_{\rm 1,th}=15\,$pA as shown in (a) at different bath temperatures $T_0=43-256\,$mK from top to bottom, against the current through JJ. (d) The equilibrium values of $V_{\rm 1,th}$, at $I=0$ (the position is shown by the red arrow in panel (c), serves as the calibration $V_{\rm 1,th} (I=0)$, versus $T$.}
		\label{fig.current-biased-meas}
	\end{figure*}

	A pair of NIS-junctions attached to each resistor $R_i$ serves as a thermometer. We apply a constant current $I_{\rm th,i}$ through this pair and measure the voltage across it~\cite{RMPJB}, see Figs.\,\ref{fig.voltage-biased-meas} and \ref{fig.current-biased-meas}. The calibration is done under equilibrium when the JJ bias current (and voltage) are set to zero. Under these conditions we assume that the electronic temperature of $R_i$ equals that of the cryostat, measured independently using a calibrated ${\rm RuO_x}$ thermometer. Figures\,\ref{fig.voltage-biased-meas} and \ref{fig.current-biased-meas} are examples of the calibrations for the data in the main text.

	\section*{Theory and fitting details}
	
	In this section we add more detail to the theory utilized to fit the experiment in the main text, and we explain how the fitting is performed. We limit ourselves to a classical description of the circuit, and neglect effects of finite temperature and electrical noise, which we expect to play only a minor role. 
	
	\begin{figure*}
		\centering
		\includegraphics [width=0.5\textwidth] {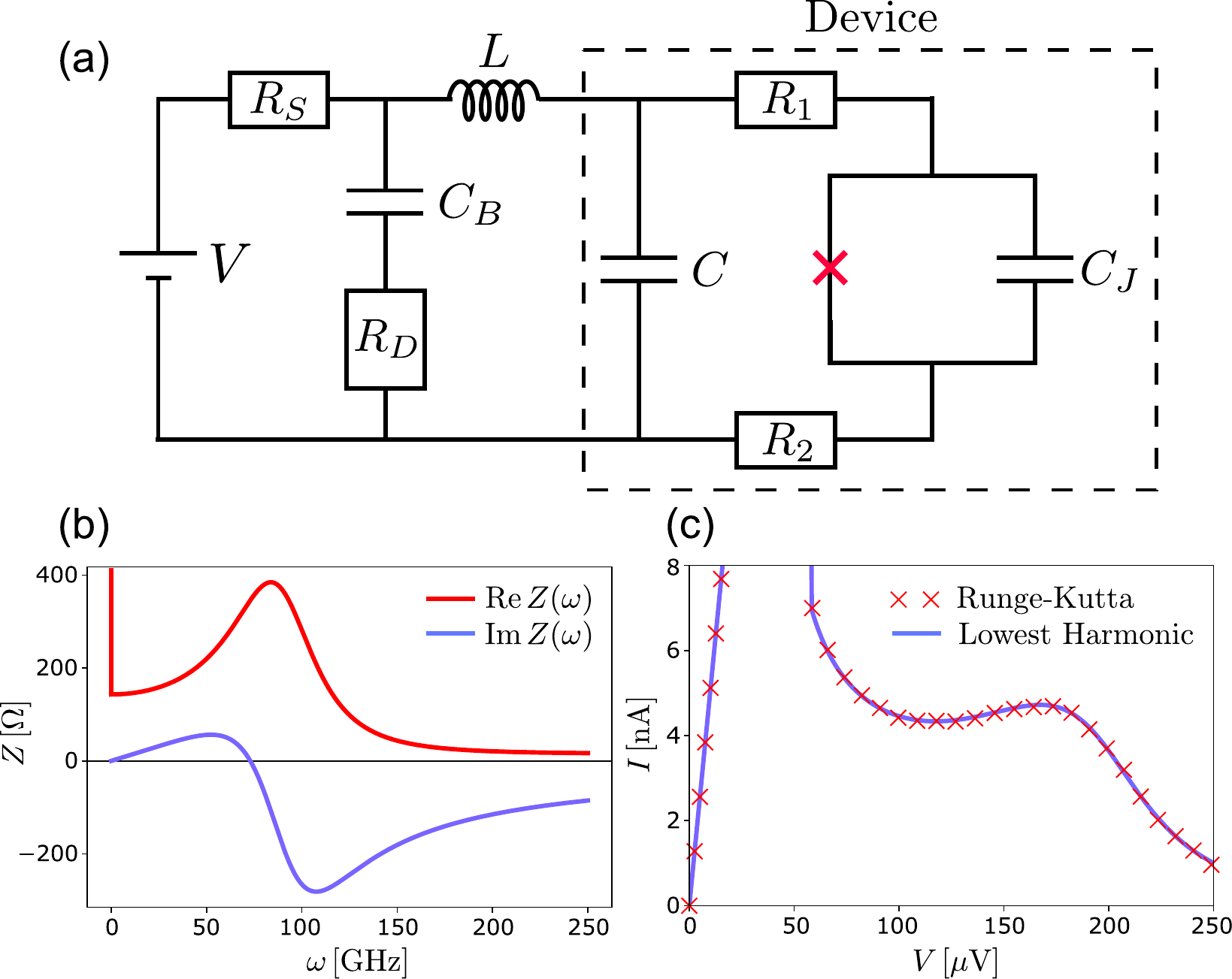}
		\caption{(a) Lump element circuit used to model the experiment. The dashed section highlights the part of the circuit stemming from the device itself. (b) Impedance for the circuit in (a) using the fit parameters from the main text: $R_S=1920$~$\Omega$, $R_1+R_2 = 30$~$\Omega$, $R_D = 120$~$\Omega$, $C = 40$~fF, $C_J = 15$~fF, $C_B \rightarrow \infty$, $L = 2.1$~nH, $I_c = 64$~nA, $V_{sw} = 0.46 Z(0)I_c$. (c) Theoretical {\it I-V} using parameters from (b). 'Runge-Kutta' is via solving Eq.~(\ref{eq:Full}) numerically, while 'Lowest Harmonic' is using Eq.~(\ref{eq:Simplified}). }
		\label{fig1sm}
	\end{figure*}
	
	As in the main text we describe the experimental circuit via lump elements, with the full circuit shown in Fig.~\ref{fig1sm}(a) with the dashed area indicating elements associated with the device, with the rest associated to the larger circuit external to the chip. Here $C_B$, which describes the contacts on the sample stage outside the chip, is assumed to be very large, $C_B\rightarrow \infty$, which separates the impedance into a zero-frequency part, $Z(\omega = 0) = R_1 + R_2 + R_S$, and a finite frequency part as
	\begin{align}
		Z(\omega) &= \left[i C_J\omega + \frac{1}{R_1+ R_2 + Z_{LC}(\omega)}\right]^{-1},
		\\
		Z_{LC}(\omega) &= \left[iC\omega + \frac{1}{iL\omega+\frac{R_S R_D}{R_S+R_D}}\right]^{-1}
	\end{align}
	The full impedance is plotted in Fig.~\ref{fig1sm}(b), using the same parameters as the fit in the main text. The main features of this impedance is a LC resonance, $\omega_{LC} \approx \frac{1}{\sqrt{L(C+C_J)}}$, which is broadened mainly by $R_D$. Before the resonance, $\omega \ll \omega_{LC}$, the real part of the impedance saturates at $Z(\omega) \approx R_1 + R_2 + \frac{R_D R_S}{R_D + R_S}$, while above the resonance the real part goes to zero as all current runs through the capacitance's, $Z(\omega) \approx \frac{-i}{(C_J + C)\omega}$. Next, we turn our attention to the Josephson junction.
	
	The principal equations of a Josephson junction are,
	\begin{equation}
		I(t) = I_c \sin\varphi(t), \hspace{0.5cm} \frac{d\varphi}{dt} = \frac{h}{2e}V_J(t)
	\end{equation}
	with $V_J(t)$ being the junction voltage drop. Putting the junction in series with $Z(\omega)$ and a source at voltage $V$ yields
	\begin{equation}
		V = V_J(t) + \int d\tau Z_t(t-\tau)I_c\sin\varphi(\tau). \label{eq:Full}
	\end{equation}
	with $Z_t(t)$ being the Fourier transform of $Z(\omega)$. For $|V| \leq Z(0)I_c$ a trivial solution exists with a constant $\sin\varphi = \frac{V}{I_c Z(0)}$, corresponding to the supercurrent branch. However, since $Z(0)>\text{Re}\hspace{0.1cm}Z(\omega)$ the circuit is underdamped with the possibility of multiple solutions. In a full treatment, finite temperature and noise would render parts of the supercurrent branch unstable with a switching voltage $V_{sw}\leq Z(0)I_c$. Here to simplify, we choose $V_{sw} = 0.46 Z(0)I_c$ to match experiment and only consider solutions for $\varphi \neq$ const. at $|V|>V_{sw}$.
	
	To obtain steady-state solutions after the switching one has to solve Eq.~(\ref{eq:Full}). This can be done numerically by Runge-Kutta method, shown in Fig.~\ref{fig1sm}(c), or analytically by keeping only the lowest harmonic of the Josephson frequency, $\omega_J/2\pi$, for which the phase is assumed to evolve as,
	\begin{equation}
		\varphi(t) = \omega_Jt + \varphi_A \sin\left(\omega_Jt+\delta\right), 
	\end{equation}
	in steady-state. Inserting this into Eq.~(\ref{eq:Full}), separating Fourier components, and keeping only the lowest order Bessel terms of $\sin\varphi(t)$, we find the following three equations,
	\begin{align}
		&V= \frac{h}{2e}\omega_J +Z(0)I_cJ_1(\varphi_A)\sin\delta, \\
		&\frac{J_0(\varphi_A)}{\varphi_A} = \frac{h}{2e}\frac{\omega_JI_c}{|Z(\omega_J)|}, \\
		&\delta = -\arctan\left(\frac{\text{Re}\hspace{0.05cm}Z(\omega_J)}{\text{Im}\hspace{0.05cm}Z(\omega_J)}\right).
	\end{align}
	This can be further simplified by assuming $\varphi_A$ to be small, and by Taylor expanding we find,
	\begin{equation}
		I_0 = I_c^2\frac{e}{\hbar}\frac{\text{Re}\hspace{0.05cm}Z(\omega_J)}{\omega_J}, \hspace{0.5cm}
		V = \frac{\hbar}{2e}\omega_J + Z(0)I_0, \label{eq:Simplified}
	\end{equation}
	with $I_0$ denoting DC current and matching Eqs.~(2) of the main text. These equations can be jointly solved to obtain the {\it I-V}, shown as the full line in Fig.~\ref{fig1sm}(c). The perfect correspondence between numerical solutions of Eq.~(\ref{eq:Full}) and  Eq.~(\ref{eq:Simplified}) supports this expansion. Lastly, we comment that an identical result to Eq.~(\ref{eq:Simplified}) can be obtained from $P(E)$ theory by expanding to lowest order in $\text{Re}\hspace{0.05cm}Z(\omega)/R_K$, with $R_K=\frac{h}{2e^2}$ denoting the resistance quantum, highlighting that the classical limit is justified \cite{Ingold2005Aug,Bretheau2013Feb}. To obtain the DC power deposited in HEBs we use $P_j = R_j \langle I_{j}^2\rangle_0$, with $I_j$ being the current through HEB $j$, and index $0$ indicating DC component. The full current and voltage drop across the junction is given by,
	\begin{align}
		V_J(t) &= \frac{h}{2e}\omega_J - I_c\text{Re}\hspace{0.05cm}Z(\omega_J)\sin(\omega_J t) - I_c\text{Im}\hspace{0.05cm}Z(\omega_J)\cos(\omega_Jt), \\
		I(t) &= I_0 + I_c \sin(\omega t),
	\end{align}
	using eqs.~(\ref{eq:Simplified}). The current through a HEB is then given by,
	\begin{equation}
		I_j(t) = I(t) - C_J\frac{dV_J}{dt} 
	\end{equation}
	which is used to obtain Eq.~(3) of the main text. To note, for $\omega_J \gg \omega_{LC}$ all current runs through the capacitors and $I_j(t) \approx \frac{I_c}{1 + C_J/C}\sin \omega_Jt $, hightlighting that the ratio $C_J/C$ controls the distribution of power in this regime.
	
	\begin{figure*}
		\centering
		\includegraphics [width=0.8\textwidth] {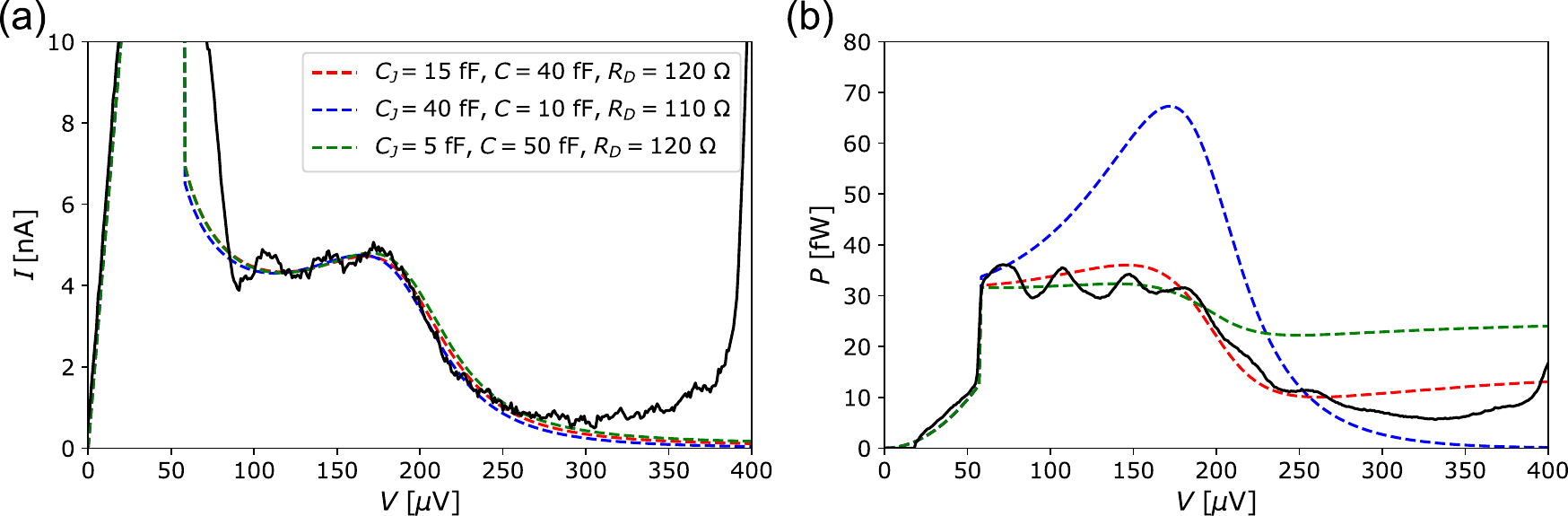}
		\caption{(a) {\it I-V}'s fitted to experiment for different values of $C_J$, $C$ and $R_D$, with remaining parameters matching those in Fig.~\ref{fig1sm}(b). (b) Power absorbed in $R_1$ for the three parameter sets in (a), highlighting the importance of the $C/C_J$ scale.} 
		\label{fig2sm}
	\end{figure*}
	
	Next, we turn to the fitting of circuit parameters used in the main paper. First, the resistance of the thermometers is estimated to be $R_1+R_2 = 30$~$\Omega$, and then from matching the supercurrent branch we find $R_S = 1920$~$\Omega$. Also, the junction capacitance with parallel line capacitance (between the junction and resistors) is geometrically estimated to be $C_J\approx 15$~fF. Next, there are two main features we utilize to extract parameters; 1) to fit the experimental {\it I-V} we require an LC frequency $\omega_{LC}/2\pi\approx 93$~GHz and $R_D \approx 120$~$\Omega$ to place and broaden the resonance correspondingly, 2) to match the drop in power in regime 3 we require $C \approx 40$~fF. These requirements constrain the inductance to $L = 2.1$~nH, thereby fixing all parameters. This value of $L$ corresponds to a few mm long bonding wire. In general, the measurement of power allows us to determine the capacitive and inductive terms more precisely. This is highlighted in Fig.~\ref{fig2sm} where we show multiple fits for the {\it I-V}'s of the main text. Here, the {\it I-V}'s are of almost equal quality, but plotting power for the same parameters reveals differences. Lastly, we highlight that the ratio of $R_D$ to $R_{j}$ determines the ratio of power absorbed by the HEBs, and the value of $R_D = 120$~$\Omega$, found by fitting {\it I-V}, fits well with the experiment, where we estimate this ratio to be $\sim 25\%$.

\end{document}